\documentclass[letterpaper,11pt,aps,prd,amsmath,amssymb,nofootinbib,reprint]{revtex4-1}

\usepackage{graphicx}
\usepackage[colorlinks=true,hyperfootnotes=false]{hyperref}
\usepackage{amsmath}
\usepackage{tabularx}
\usepackage{amsfonts}
\usepackage{amssymb}
\usepackage{dsfont}
\usepackage{graphicx,color,dcolumn,booktabs}
\usepackage{txfonts}
\usepackage{amssymb}
\newcommand{\drmd}[1]{m_{D,#1}}
\newcommand{\drmds}[1]{m_{D^\ast,#1}}
\allowdisplaybreaks
\begin{document}

\title{\boldmath $Z_c(3900)$: Confronting theory and lattice simulations}

\author{Miguel Albaladejo}
\affiliation{Instituto de F\'isica Corpuscular (IFIC), Centro Mixto CSIC-Universidad de Valencia, \\
Institutos de Investigaci\'on de Paterna, Aptdo. 22085, E-46071 Valencia, Spain}

\author{Juan Nieves}
\affiliation{Instituto de F\'isica Corpuscular (IFIC), Centro Mixto CSIC-Universidad de Valencia, \\
Institutos de Investigaci\'on de Paterna, Aptdo. 22085, E-46071 Valencia, Spain}

\author{Pedro Fernandez-Soler}
\affiliation{Instituto de F\'isica Corpuscular (IFIC), Centro Mixto CSIC-Universidad de Valencia, \\
Institutos de Investigaci\'on de Paterna, Aptdo. 22085, E-46071 Valencia, Spain}

% \date{\today}

\begin{abstract}
We consider a recent $T$-matrix analysis by Albaladejo {\it et al.}, [Phys.\ Lett.\ B {\bf 755}, 337 (2016)] which accounts for the $J/\psi\pi$ and
$D^\ast\bar{D}$ coupled--channels dynamics, and that successfully describes the
experimental information concerning the recently discovered
$Z_c(3900)^\pm$. Within such scheme, the data can be similarly well
described in two different scenarios, where the $Z_c(3900)$ is either a
resonance or a virtual state. To shed light into the nature of this
state, we apply this formalism in a finite box with the aim of
comparing with recent Lattice QCD (LQCD) simulations. We see that the energy
levels obtained for both scenarios agree well with those obtained in the
single-volume LQCD simulation reported in Prelovsek {\it et al.} [Phys.\ Rev.\ D
{\bf 91},  014504 (2015)], making thus difficult  to disentangle between
both possibilities. We also study the volume
dependence of the energy levels obtained with our formalism, and
suggest that LQCD simulations performed at several volumes could help
in discerning the actual nature of the intriguing $Z_c(3900)$ state.
\end{abstract}

\maketitle

\section{Introduction}

Since the discovery of the $X(3872)$ in 2003 \cite{Choi:2003ue}, the
charmonium and charmonium-like spectrum are being continuously enlarged
with new so-called $XYZ$ states \cite{Olsen:2014qna,Chen:2016qju,Hosaka:2016pey},
many of which do not fit properly in the conventional quark models
\cite{Godfrey:1985xj}. The relevance of meson-meson channels can be
grasped from the fact that all the charmonium states predicted below
the lowest hidden-charm threshold ($D\bar{D}$) have been
experimentally confirmed, but above this energy most of the observed
states cannot be unambiguously identified with any of the 
predicted charmonium $c \bar c$ states.

Amongst the $XYZ$ states, the $Z_c(3900)^\pm$ was simultaneously
discovered by the BESIII and Belle collaborations
\cite{Ablikim:2013mio, Liu:2013dau} in the $e^+ e^- \to Y(4260) \to
J/\psi\pi^+ \pi^-$ reaction, where a clear peak very close to the $D^\ast
\bar{D}$ threshold, around $3.9\ \text{GeV}$, is seen in the $J/\psi
\pi$ spectrum. Later on, an analysis \cite{Xiao:2013iha} based on
CLEO-c data for a different reaction, $e^+e^- \to \psi(4160) \to
J/\psi\pi^+\pi^-$, confirmed the presence of this resonant structure
as well, although with a somewhat lower mass. The BESIII collaboration
\cite{Ablikim:2013xfr, Ablikim:2015swa} has also reported a
resonant-like structure in the $\bar{D}^\ast D$ spectrum for the
reaction $e^+ e^-\to \bar D^* D\pi$ at different $e^+e^-$
center-of-mass (c.m.) energies [including the production of
  $Y(4260)$]. This structure, with quantum numbers favored to be
$J^P=1^+$, has been cautiously called $Z_c(3885)^\pm$, because its fitted
mass and width showed some differences with those attributed to the
$Z_c(3900)^\pm$.  Whether both set of observations correspond to the 
same state needs to be confirmed, though there is a certain
consensus that this is indeed the case, and the peaks reported as the
$Z_c(3885)^\pm$ and $Z_c(3900)^\pm$ are originated by the same state
seen in different channels. Moreover, evidence for its neutral partner,
$Z_c(3900)^0$, has also been reported
\cite{Xiao:2013iha,Ablikim:2015tbp}.

The  nature of the $Z_c\left(3900\right)^{\pm}$ is
intriguing. On one hand, it couples to $D^\ast \bar{D}$ and $J/\psi
\pi$, and therefore one assumes it should contain a constituent
$c\bar{c}$ quark--anti-quark pair. On the other hand, it is charged
and hence it must also have another constituent quark--anti-quark
pair, namely $u\bar{d}$ (for $Z_c^+$). Its minimal structure would be
then $c\bar{c}u\bar{d}$, which automatically qualifies it as a
non-$q\bar{q}$ (exotic) meson. Being a candidate for an exotic hidden
charm state, it has triggered much theoretical interest. An early
discussion of possible structures for  the
$Z_c\left(3900\right)^{\pm}$ was given in
Ref.~\cite{Voloshin:2013dpa}. The suggested interpretations cover a
wide range: a $\bar{D}^\ast D$
molecule~\cite{Wang:2013cya,Guo:2013sya,Wilbring:2013cha,Dong:2013iqa,Zhang:2013aoa,Ke:2013gia,Aceti:2014uea,He:2015mja},
a tetraquark
\cite{Braaten:2013boa,Wang:2013vex,Dias:2013xfa,Deng:2014gqa,Qiao:2013raa,Esposito:2014rxa,Maiani:2014aja},
an object originated from an attractive $\bar{D}^\ast D^\ast$
interaction \cite{Zhou:2015jta}, a simple kinematical
effect~\cite{Chen:2013coa,Swanson:2014tra}, a cusp enhancement due to
a triangle singularity \cite{Szczepaniak:2015eza}, or a radially
excited axial meson \cite{Coito:2016ads}. In Ref.~\cite{Guo:2014iya},
it was argued that this structure cannot be a kinematical effect
and that it must necessarily be originated from a nearby
pole. Consequences from some of these models have been 
discussed in Ref.~\cite{Cleven:2015era}. The non-compatibility
(partial or total) of the properties of the $Z_c$ deduced in 
different  approaches clearly hints why 
 the actual nature of this state has attracted so much
attention.

In Ref.~\cite{Albaladejo:2015lob},  theoretical basis of
the present manuscript, a $J/\psi \pi$--$D^\ast \bar{D}$
coupled-channels scheme was proposed to describe the observed
peaks associated to the $Z_c(3900)$, which is assumed to have
$I(J^{PC})=1(1^{+-})$  quantum numbers.\footnote{Through all this work,
  charge conjugation refers only to the neutral element of the $Z_c(3900)$ isotriplet.} Within this coupled channel
scheme, it was possible to  successfully describe simultaneously the BESIII $J\psi \pi$
\cite{Ablikim:2013mio} and $D^\ast \bar{D}$ \cite{Ablikim:2015swa}
invariant mass spectra, in which the $Z_c(3900)^\pm$ structure has been
seen. Interestingly, two different fits with similar quality were able
to reproduce the data. In each of them, the origin of the
$Z_c(3900)^\pm$ was different. In the first scenario, it corresponded
to a resonance originated from a pole above the $D^\ast \bar{D}$
threshold, whereas in the second one the structure was produced by
a virtual pole below the threshold (see Ref.~\cite{Albaladejo:2015lob}
for more details).

Hadron interactions are governed by the non-perturbative regime of QCD
and, for this reason, Lattice QCD (LQCD) is an essential theoretical
tool in hadron physics. In particular, one of the aims of LQCD is to
obtain the hadron spectrum from quarks and gluons and their
interactions (see {\it e.g.} Ref.\cite{Fodor:2012gf}
for a review focused on the light sector, and
Refs.~\cite{Dudek:2007wv,Bali:2011rd,Liu:2012ze,Lang:2015sba} for
results concerning the charmonium sector). For such a purpose the L\"uscher method
\cite{Luscher:1986pf,Luscher:1990ux} is widely used. It relates the discrete energy
levels of a two-hadron system in a finite box with the phase shifts
and/or binding energies of that system in an infinite
volume. Appropriate generalizations relevant for our work can be found
in Refs.~\cite{Liu:2005kr,Lage:2009zv,Bernard:2010fp,Doring:2011vk}.

LQCD simulations devoted to find the $Z_c(3900)$ state are still scarce
\cite{Prelovsek:2013xba,Prelovsek:2014swa,Chen:2014afa,Liu:2014mfy,Lee:2014uta,Ikeda:2016zwx}. Exploratory
theoretical studies for hidden charm molecules have been performed in
Refs.~\cite{Albaladejo:2013aka,Garzon:2013uwa}, while actual LQCD simulations~\cite{Prelovsek:2013xba,Prelovsek:2014swa,Chen:2014afa,Liu:2014mfy,Lee:2014uta}
find energy levels showing a weak interaction in the $Z_c(3900)^\pm$
quantum-numbers  sector (either attractive or
repulsive), and no evidence is found for its existence. 
The work of Ref.~\cite{Ikeda:2016zwx} employs LQCD to obtain
a coupled-channel $S$-matrix, which shows an interaction dominated by
off-diagonal terms, and, according to Ref.~\cite{Ikeda:2016zwx}, this
does not support a usual resonance picture for the $Z_c(3900)$. This
$S$-matrix contains a pole located well below threshold in an
unphysical Riemann sheet, {\it i.e.}, a virtual pole. It is worth to
note that this possibility could be in agreement with the second
scenario advocated in Ref.~\cite{Albaladejo:2015lob}, and mentioned above.

Our objective in the present manuscript is to implement the coupled
channel $T$-matrix fitted to data in Ref.~\cite{Albaladejo:2015lob} in
a finite volume and study its spectrum. Thus, we will be able to
compare the energy levels obtained with this finite volume $T$-matrix  with
those obtained in LQCD simulations, in particular those reported in
Ref.~\cite{Prelovsek:2014swa}. This work is organized as follows. The
formalism is presented in Sec.~\ref{sec:form}, while the $T$-matrix of
Ref.~\cite{Albaladejo:2015lob} is briefly discussed in
Subsec.~\ref{subsec:form_iv}, and its extension for a finite volume is
outlined in Subsec.~\ref{subsec:form_fv}. Results are presented and
discussed in Sec.~\ref{sec:res}, and the conclusions of this work,
together with  a brief summary are
given in Sec.~\ref{sec:sum}.

\section{Formalism}\label{sec:form}
\begin{table*}[tbh]
\caption{Values of the parameters employed in Eq.~\eqref{Eq:C-matrix},
  taken from Ref.~\cite{Albaladejo:2015lob}, together with the $Z_c$ pole
  positions found in that work. The errors account for statistical
  (first) and systematic (second) uncertainties (see
  Ref.~\cite{Albaladejo:2015lob} for details).\label{Table:C-parameters}}
\begin{tabular}{ccccccc}  \hline\hline
\vphantom{$\displaystyle \frac{a}{b}$} $\Lambda_2\ (\text{GeV})$ & $C_{1Z}\ (\text{fm}^2)$ & $b\ (\text{fm}^3)$ & $\widetilde{C}\ (\text{fm}^2)$ & $M_{Z_c}\ (\text{MeV})$ & $\Gamma_{Z_c}/2\ (\text{MeV})$ \\ \hline\vspace{0.1cm}
$1.0$ & $          - 0.19 \pm 0.08 \pm 0.01$ & $ -2.0 \pm 0.7 \pm 0.4$ & $0.39 \pm 0.10 \pm 0.02$ & $3894 \pm 6 \pm 1$                                & $30 \pm            12 \pm 6$  \\\vspace{0.1cm}
$0.5$ & $\hphantom{+}0.01 \pm 0.21 \pm 0.03$ & $ -7.0 \pm 0.4 \pm 1.4$ & $0.64 \pm 0.16 \pm 0.02$ & $3886 \pm 4 \pm 1$                                & $22 \pm \hphantom{1}6 \pm 4$  \\\vspace{0.1cm}
$1.0$ & $          - 0.27 \pm 0.08 \pm 0.07$ & $0$ (fixed)             & $0.34 \pm 0.14 \pm 0.01$ & $3831 \pm 26^{+\hphantom{1}7}_{-28}$              & virtual state                 \\\vspace{0.1cm}
$0.5$ & $          - 0.27 \pm 0.16 \pm 0.13$ & $0$ (fixed)             & $0.54 \pm 0.16 \pm 0.02$ & $3844 \pm 19^{+           12}_{-21}$              & virtual state                 \\ \hline \hline
\end{tabular}
\end{table*}
\subsection{Infinite volume}\label{subsec:form_iv}
We first briefly review the model of Ref.~\cite{Albaladejo:2015lob} (where the
reader is referred for more details) that we are going to employ
here. There,  the $Y(4260)$  decays to $D\bar{D}^{\ast}\pi$ and
$J/\psi \pi\pi$ are studied with a model shown diagrammatically in 
Fig. 1 of that reference.  Final state interactions among the outgoing
$D\bar{D}^{\ast}$ and $J/\psi \pi$ produce  the peaks observed by the
BESIII collaboration, which are associated to  the $Z_c(3900)$ state. 
The two channels involved in the $1(1^{+})$  $T$-matrix are denoted   as
$1\equiv J/\psi \pi$ and $2\equiv D\bar{D}^{\ast}$. Solving the
on-shell version of the factorized Bethe-Salpeter equation (BSE) allows to
write:
\begin{align}
 T^{-1}(E)=V^{-1}(E) - G(E),
 \label{Eq:T-matrix}
\end{align}
where $E$ is the c.m. energy of the system. The symmetric $V$ matrix is the potential kernel, whose matrix elements have the following form:
\begin{align}
 V_{ij}=4\sqrt{m_{i,1}m_{i,2}m_{j,1}m_{j,2}}\ C_{ij}\ e^{-k^2_i/\Lambda^2_i}e^{-k^2_j\,/\Lambda^2_j}.
 \label{Eq:potential}
\end{align}
with $m_{i,1}$ and $m_{i,2}$  the masses of the particles of the $i$th
channel and $k_i^2$, the relative three-momenta squared in the
c.m. frame, implicitly defined through:
\begin{align}
E & = \omega_\psi(k_1) + \omega_\pi(k_1), \label{eq:mom1} \\
E & = \omega_{D^\ast\bar{D}}(k_2)~, \label{eq:mom2}
\end{align}
where:
\begin{align}
\omega_\psi(q) & = \sqrt{m_{J/\psi}^2+q^2}~,\label{eq:om_psi}\\
\omega_\pi(q)  & = \sqrt{m_{\pi}^2+q^2}~,   \label{eq:om_pi}\\
\omega_{D^\ast\bar{D}}(q) & = m_D + m_{D^\ast} + \frac{m_D+m_{D^\ast}}{2 m_D m_{D^\ast}} q^2~.\label{eq:omDD}
\end{align}
with $q \equiv |\vec{q}\,|$. The Gaussian form factors
$e^{-k^2_i/\Lambda^2_i}$ are introduced to regularize the BSE, and thus,
for each channel, an ultraviolet (UV) cut-off $\Lambda_i$ is
introduced. In this work, we have used $\Lambda_1 = 1.5\ \text{GeV}$ and two values for
$\Lambda_2 = 0.5$ and 1
GeV~\cite{Nieves:2012tt,HidalgoDuque:2012pq}. The $C_{ij}$ matrix
stands for the $S$-wave interaction in the coupled-channels space, and
it is given by~\cite{Albaladejo:2015lob}:
\begin{align}
 C&=
    \begin{bmatrix}
    0 & \tilde{C} \\
    \tilde{C} & C_{22}\left(E\right) \\
    \end{bmatrix} 
 \label{Eq:C-matrix}.
\end{align}
In Eq. \eqref{Eq:C-matrix} the $J/\psi\pi\to J/\psi\pi$ interaction is
neglected, $C_{11}=0$, the inelastic transition one is approximated by
a constant, $\tilde{C}$, while the $D^\ast \bar{D} \to D^\ast \bar{D}$
potential $C_{22}(E)$ is parametrized as:
\begin{align}
 C_{22}(E)=C_{1Z}+b\left(E-m_D-m_{D^\ast}\right).
 \label{Eq:C22}
\end{align}
In a momentum expansion, the lowest order contact potential for this
elastic transition would be simply a constant, $C_{22} \equiv
C_{1Z}$. However, it is easy to prove that two coupled channels with
contact potentials cannot generate a resonance above threshold. Thus
and for the sake of generality, the model of
Ref.~\cite{Albaladejo:2015lob} allows for an energy dependence in
Eq. \eqref{Eq:C22}, driven by the $b$ parameter. The $G$ matrix in
Eq. \eqref{Eq:T-matrix} is diagonal, and its matrix elements are the 
$J/\psi \pi$ and $D^\ast \bar{D}$  loop functions,
\begin{align}
 G_{11}(E) & =\int_{\mathbb{R}^3}\frac{{\rm d}^3q}{\left(2\pi\right)^3}\frac{\omega_\psi(q)+\omega_\pi(q)}{2\omega_\psi(q)\omega_\pi(q)}\frac{e^{-2\left(q^2-k_1^2\right)/\Lambda_1^2}}{E^2-\left(\omega_\psi(q)+\omega_\pi(q)\right)^2+i\epsilon}~, \label{Eq:G-11}\\
 G_{22}(E) & =\frac{1}{4 m_D m_{D^\ast}}\int_{\mathbb{R}^3}\frac{{\rm d}^3q}{\left(2\pi\right)^3}\frac{e^{-2\left(q^2-k_2^2\right)/\Lambda_2^2}}{E-\omega_{D\bar{D}^\ast}(q)+i\epsilon}~,\label{Eq:G-22}
\end{align}
which account for the right-hand cut of the $T$-matrix, that
satisfies in this way the optical theorem. The $D^\ast \bar{D}$ channel
loop function $G_{22}$ is computed in the non-relativistic
approximation. 

The free parameters in the interaction matrix $C$ ($\tilde{C}$,
$C_{1Z}$ and $b$) were fitted in Ref.~\cite{Albaladejo:2015lob} to the
experimental $J/\psi\pi^-$ and $D^+D^{\ast -}$ invariant mass
distributions in the $Y(4260)\to J/\psi \pi\pi$ and $Y(4260)\to
D\bar{D}^\ast \pi$ decays \cite{Ablikim:2013mio,Ablikim:2015swa}. The
fitted parameters are compiled here in Table~\ref{Table:C-parameters},
where we can see the two different scenarios investigated in Ref.~\cite{Albaladejo:2015lob}. In the first one, $b\neq 0$, the
$Z_c$ appears as a $D^\ast \bar{D}$ resonance, {\it i.e.}, a pole
above the $D^\ast \bar{D}$ threshold in a Riemann sheet connected with
the physical one above this energy. In the second one, where $b=0$, a
pole appeared below the $D\bar{D}^\ast$ threshold in an unphysical
Riemann sheet, which gives rise to the $Z_c(3900)$ structure, peaking
exactly at the $D^\ast \bar{D}$ threshold in this case
\cite{Albaladejo:2015lob} (see also Ref.~\cite{Albaladejo:2015dsa}).

\subsection{Finite volume}\label{subsec:form_fv}
In this subsection, we study the previous coupled channel $T$-matrix
in a finite volume. The consequence of putting the interaction in a
box of size $L$ with periodic boundary conditions is that the
three-momentum is no longer a continuous variable, but a discrete
one. For each value of $L$, we have the infinite set of momenta
$\vec{q}=\frac{2\pi}{L} \vec{n}$, $\vec{n} \in\mathbb{Z}^3$. The integrals in Eqs. \eqref{Eq:G-11}
and \eqref{Eq:G-22} will be replaced by  sums over all
the possible values of $\vec{q}$:
\begin{align}
 \tilde{G}_{11}(E)& =\frac{1}{L^3}\sum_{\vec{n}}\frac{\omega_\psi(q)+\omega_\pi(q)}{2\omega_\psi(q)\omega_\pi(q)}\frac{e^{-2\left(q^2-k_1^2\right)/\Lambda_1^2}}{E^2-\left(\omega_\psi(q)+\omega_\pi(q)\right)^2}~,
 \label{Eq:G-11-FV}\\
 \tilde{G}_{22}(E)& =\frac{1}{4\left(m_Dm_{D^\ast}\right)}\frac{1}{L^3}\sum_{\vec{n}}\frac{e^{-2\left(q^2-k_2^2\right)/\Lambda_2^2}}{E-\omega_{D\bar{D}^\ast}(q)}~,
 \label{Eq:G-22-FV}
\end{align}
(see Ref.~\cite{Albaladejo:2013aka} for further details). The $T$-matrix
in a finite volume is then:
\begin{equation}\label{eq:ttilde}
\tilde{T}^{-1}(E) = V^{-1}(E) - \tilde{G}(E)~,
\end{equation}
where the $\tilde{G}$ matrix elements are given by
Eqs.~\eqref{Eq:G-11-FV} and \eqref{Eq:G-22-FV}. The discrete energy
levels in the finite box are given by the poles of the
$\tilde{T}$-matrix. If the interaction is switched off, $V \to 0$, the
free (or non-interacting) energy levels are given by the poles of the
$\tilde{G}_{ii}$ functions,
\begin{align}
E_{J/\psi \pi}^{(\vec{n}\,^2)}     & = \omega_\psi(q_L n) + \omega_\pi(q_L n)~, \label{eq:free-1}\\
E_{D^\ast \bar{D}}^{(\vec{n}\,^2)} & = \omega_{D\bar{D}^\ast}(q_L n)~,\label{eq:free-2}
\end{align}
where we use the shorthand $q_L = 2\pi/L$, and $n = \sqrt{\vec{n}\,^2}$. The effect of the interaction is to shift these non-interacting energy levels.

Our purpose is to make contact with the results reported in the LQCD simulation of
Ref.~\cite{Prelovsek:2014swa}, and hence we will employ the masses and
the energy-momentum dispersion relations used in that work. For the
$J/\psi \pi$ channel the dispersion relation in Eq.~\eqref{eq:mom1} is
still appropriate, but for the case of the $D^\ast \bar{D}$ channel, in
Eqs.~\eqref{eq:mom2}  and \eqref{eq:omDD}, $\omega_{D\bar{D}^\ast}(q)$ must be replaced
by~\cite{Prelovsek:2014swa,Lang:2014yfa}:
\begin{equation}\label{Eq:dispersion-22}
\omega^\text{lat}_{D\bar{D}^\ast}(q)=\drmd{1} + \drmds{1} + \frac{\drmd{2}+\drmds{2}}{2\drmd{2}\drmds{2}}q^2-\frac{\drmd{4}^3+\drmds{4}^3}{8\drmd{4}^3\drmds{4}^3} q^4~.
\end{equation}
This lattice energy of the $D^\ast \bar{D}$ pair suffers from
discretization errors and it must be used in Eq.~\eqref{Eq:G-22-FV}. The
non-interacting energy levels in Eq.~\eqref{eq:free-2} should be also
modified accordingly. Notice that, because of the factor
$e^{-q^2/\Lambda^2}$, the sum in Eq.~\eqref{Eq:G-22-FV} is
exponentially suppressed in $\vec{n}\,^2$. For the range of energies
considered in this work, it is sufficient to add terms up to $\vec{n}\,^2 =
6$.\footnote{We have checked that the numerical differences are
  negligible if larger values, say $\vec{n}\,^2 = 8$, are used.}
\begin{table}[t]
\caption{Lattice parameters taken from
  Refs.~\cite{Prelovsek:2014swa,Lang:2014yfa}, and employed in this work.\label{Table:Lattice-parameters}}
\begin{tabularx}{0.3\textwidth}{p{0.07\textwidth} c l p{0.07\textwidth}}
\hline\hline
\multicolumn{4}{c}{Lengths (fm)} \\ \hline
& $a$     & $0.1239(13)$& \\ \vspace{0.1cm} 
& $L=16a$ & $1.982(21)$ & \\
\\ 
\multicolumn{4}{c}{Masses (lattice units)} \\ \hline \vspace{0.2cm}
& $am_\pi$       & $0.1673(16)$ & \\ \vspace{0.1cm} 
& $am_{J/\psi}$  & $1.54171(43)$& \\ \vspace{0.1cm} 
& $am_{\eta_c}$  & $1.47392(31)$& \\ \vspace{0.1cm} 
& $a\drmd{1}$    & $0.9801(10) $& \\ \vspace{0.1cm} 
& $a\drmd{2}$    & $1.107(12) $& \\ \vspace{0.1cm} 
& $a\drmd{4}$    & $1.107(27) $& \\ \vspace{0.1cm} 
& $a\drmds{1}$   & $1.0629(13)$& \\ \vspace{0.1cm} 
& $a\drmds{2}$   & $1.267(21)$& \\ \vspace{0.1cm} 
& $a\drmds{4}$   & $1.325(68)$& \\ 
\hline\hline
\end{tabularx}
\end{table}
Finally, the discrete, interacting energy levels reported in Ref.~\cite{Prelovsek:2014swa} are actually the result of applying the following shift:
\begin{equation}\label{eq:shift}
E \to E^\ast = E - m^\text{lat}_\text{s.a.} + m^\text{exp}_\text{s.a.}~,
\end{equation}
where the spin-average mass $m_\text{s.a.}$ is given by $m_\text{s.a.}
= \frac{1}{4}(m_{\eta_c} + 3 m_{J/\psi})$. For this reason, we will
also present our energy levels shifted as in Eq.~\eqref{eq:shift}. The
parameters involved in our calculations, taken from
Refs.~\cite{Prelovsek:2014swa,Lang:2014yfa}, are collected in Table
\ref{Table:Lattice-parameters}. In particular, one has $m_\pi = 266
\pm 4\ \text{MeV}$ and $L = 16\,a = 1.98 \pm 0.02\ \text{fm}$, being $a$ the lattice spacing.

\subsection{Further comments}\label{subsec:comm}
With all the ingredients presented in Subsec.~\ref{subsec:form_fv}, we
can compare our predictions for the energy levels in a box with those
reported in Ref.~\cite{Prelovsek:2014swa}. But before presenting our
results we would like to discuss some technical details concerning two
differences that could affect the comparison.

First,  we would like to note  that  the LQCD simulation
in Ref.~\cite{Prelovsek:2014swa} includes the $J/\psi \pi$ and $D^\ast \bar{D}$
channels that are present in our $T$-matrix analysis, but it also
includes other channels (like $\eta_c \rho$ or $D^\ast
\bar{D}^\ast$). However, according to Ref.~\cite{Albaladejo:2015lob},
it is sufficient to include the $J/\psi\pi$
and $D^\ast \bar{D}$ channels to achieve a good reproduction of the
experimental information concerning the $Z_c(3900)$. For this reason,
we expect that, in first approximation, these other
channels could be safely  neglected in the calculations.

The second point to be noted is that we are ignoring the possible
$m_\pi$ dependence of the parameters in the potential,
Eq.~\eqref{Eq:C-matrix}. Nonetheless, the LQCD simulation of
Ref.~\cite{Prelovsek:2014swa} is performed for a relatively low pion
mass, $m_\pi = 266 \pm 4\ \text{MeV}$, and we thus expect the
eventual dependence to be mild. Furthermore, we are going to compare
several sets of these parameters (presented in
Table~\ref{Table:C-parameters}), which somewhat compensates this
effect.

\section{Results and discussion}\label{sec:res}
\begin{figure*}[t]\centering
 \includegraphics[height=5.0cm,keepaspectratio]{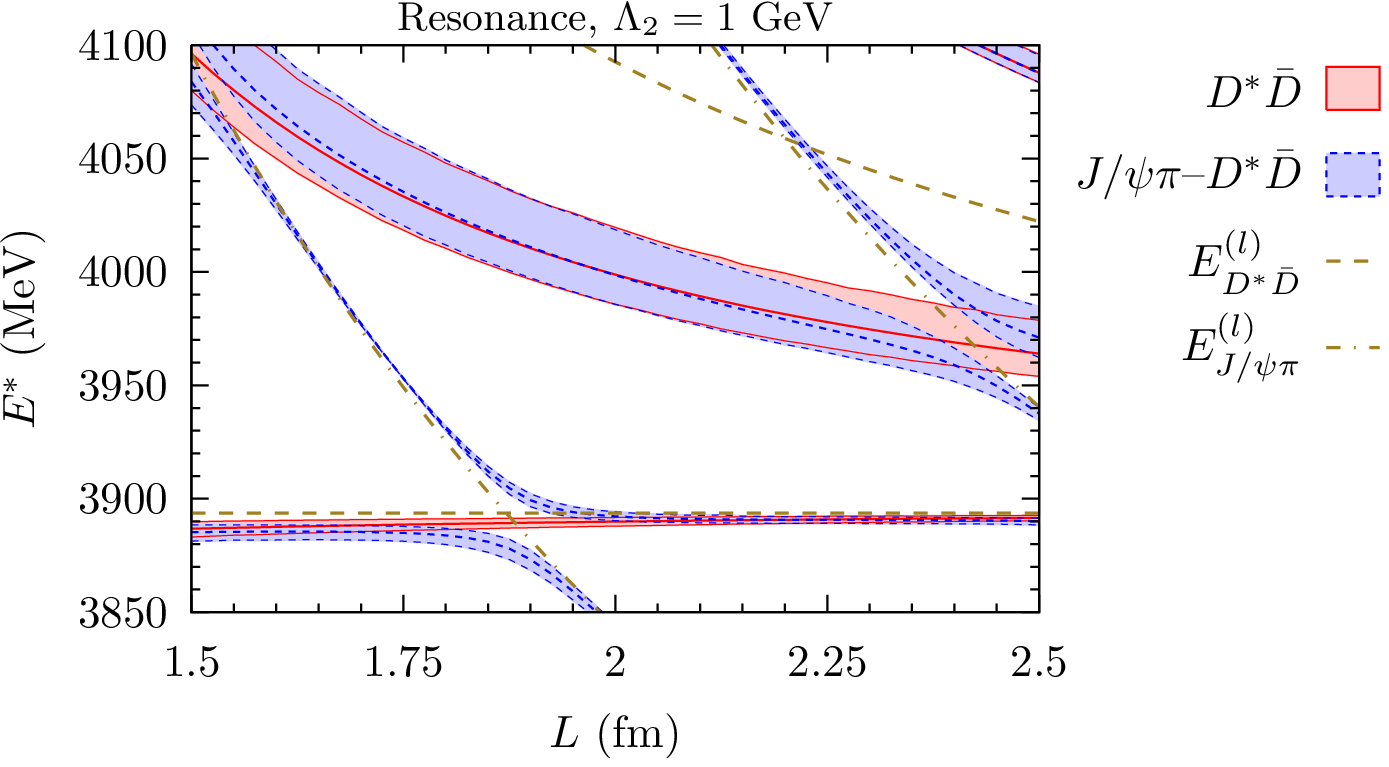} \hspace{0.1cm}
 \includegraphics[height=5.0cm,keepaspectratio]{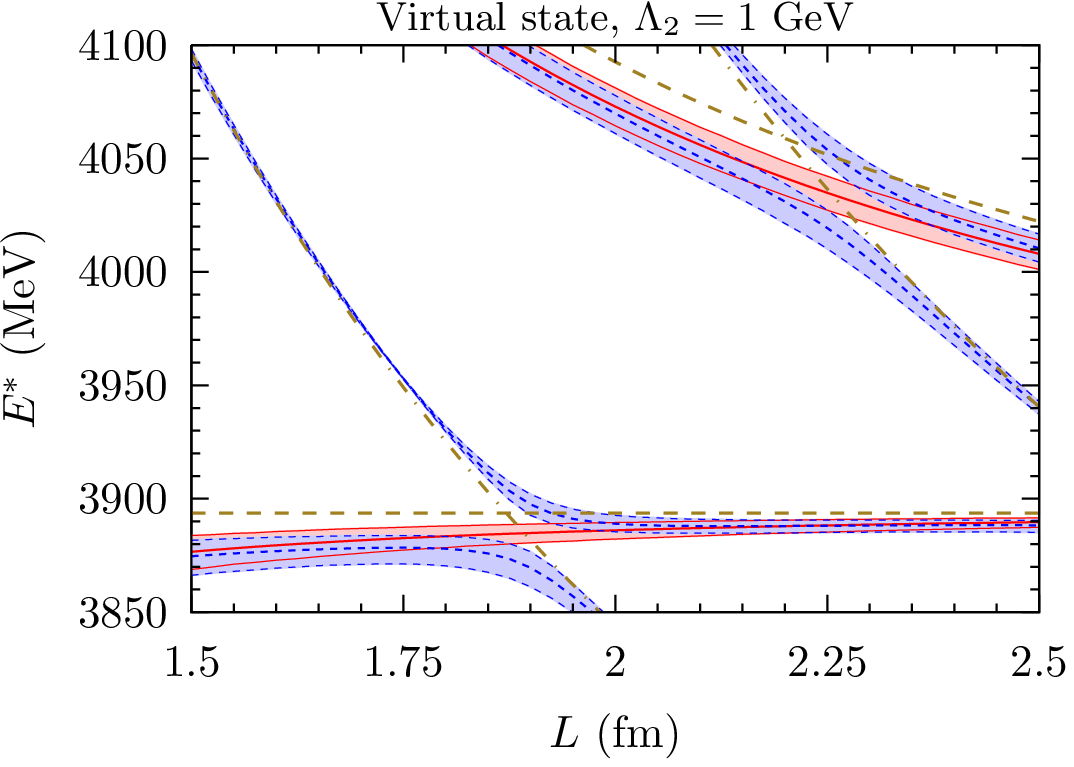} 
 \caption{Volume dependence of some energy levels located close to the
   $D^\ast \bar{D}$ threshold, and obtained when the $Z_c$ is
   described as a resonance (left) or as a virtual state (right) in the
   $L\to \infty$ limit. The blue dashed lines have been obtained from
   the $J/\psi \pi$--$D^\ast \bar{D}$ coupled channel analysis, and
   the red solid lines show the single elastic channel
   ($D^\ast\bar{D}$) case, in both cases $\Lambda_2$ has been fixed to
   1 GeV. The error bands are obtained from the uncertainties
   of the parameters introduced in the theoretical model of
   Ref.~\cite{Albaladejo:2015lob} (Table~\ref{Table:C-parameters}),
   adding in quadratures the statistical and systematic errors. The
   green dashed (dotted-dashed) lines are the free $D^\ast\bar{D}$
   ($J/\psi\pi$) energy levels $E^{(l\,)}_{D^\ast\bar{D}}$
   ($E^{(l\,)}_{J/\psi\pi}$).}
 \label{fig:energy_levels}
\end{figure*}
\begin{figure*}[t]\centering
 \includegraphics[height=5.0cm,keepaspectratio]{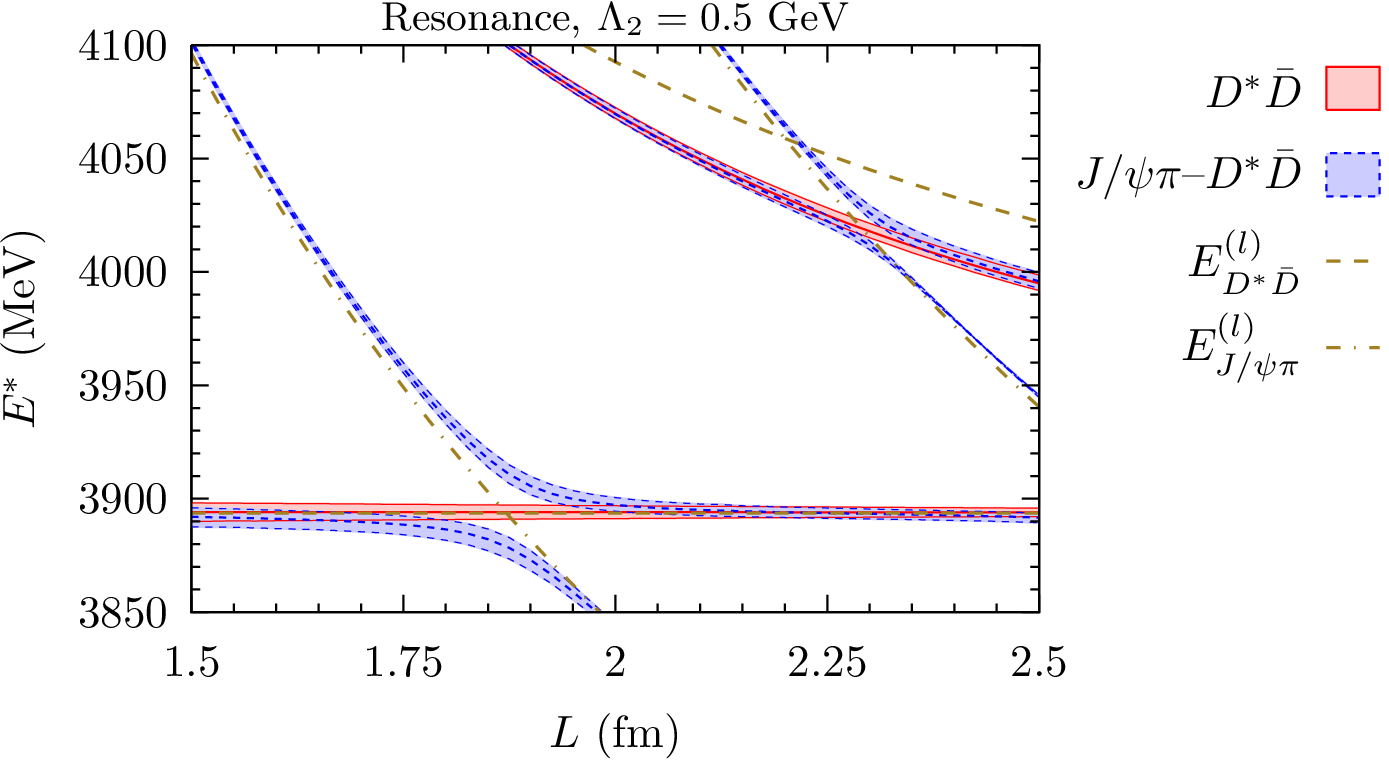} \hspace{0.1cm}
 \includegraphics[height=5.0cm,keepaspectratio]{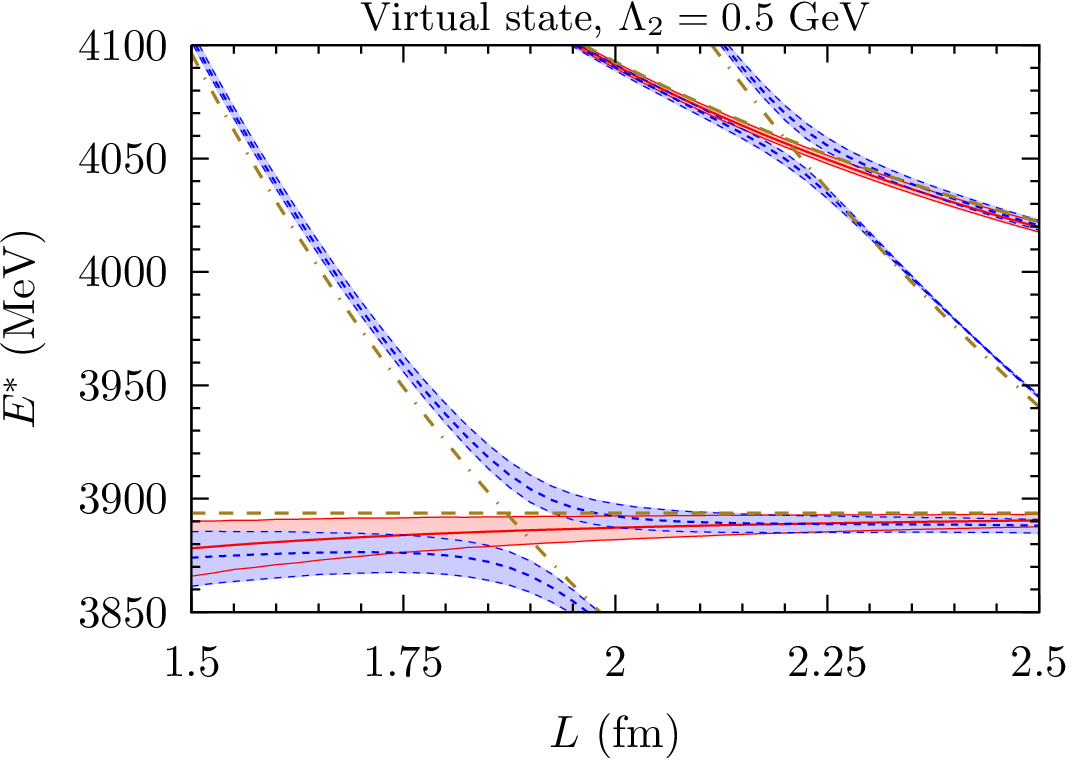} 
 \caption{Same as in Fig.~\ref{fig:energy_levels}, but for the case $\Lambda_2 = 0.5\ \text{GeV}$.}
 \label{fig:energy_levels-500}
\end{figure*}
In Fig.~\ref{fig:energy_levels}, we show the $L$ dependence of some
 energy levels close to the $D^\ast \bar{D}$ threshold. They have
 been computed from the poles of the finite volume
$\tilde{T}$-matrix, Eq.~\eqref{eq:ttilde}, by using the parameters of
Table~\ref{Table:C-parameters} for $\Lambda_2 = 1\ \text{GeV}$, and
the lattice setup given in Table~\ref{Table:Lattice-parameters}.  The
levels obtained in the $Z_c(3900)^\pm$ resonance (virtual) scenario,
calculated using the entries of the first (third) row
of Table~\ref{Table:C-parameters}, are displayed in the left (right) panel. The
blue dashed lines stand for the $J/\psi \pi$--$D^\ast \bar{D}$ 
coupled-channel-analysis results, and the red solid lines show the energy
levels obtained when the inelastic $J/\psi \pi$--$D^\ast \bar{D}$
transition is neglected ($\tilde{C}=0$). This latter case corresponds
to consider a single, elastic channel ($D^\ast\bar{D}$). The error
bands account for the uncertainties on the energy levels inherited
from the errors in the parameters of Ref.~\cite{Albaladejo:2015lob},
quoted in Table~\ref{Table:C-parameters} (statistical and systematical
errors are added in quadrature for the calculations). The green dashed
(dotted-dashed) lines stand for the non-interacting $D^\ast\bar{D}$
($J/\psi\pi$) energy levels. In Fig.~\ref{fig:energy_levels-500}, the
same results are shown but for the case $\Lambda_2 =
0.5\ \text{GeV}$. The qualitative $L$ behavior of both
Figs.~\ref{fig:energy_levels} and \ref{fig:energy_levels-500} is
similar, so we discuss first Fig.~\ref{fig:energy_levels} and, later
on, the specific differences between them will be outlined.

For both resonant and virtual scenarios, there is always an energy
level very close to a free energy of the $J/\psi\pi$  state,
$E^{(l\,)}_{J/\psi\pi}$, which reveals that the interaction driven by this
meson pair is weak. Furthermore, the energy levels for the
coupled--channel $\tilde{T}$-matrix basically follow those obtained
within the elastic $D^\ast \bar{D}$ approximation, except in the neighborhood of
the $J/\psi\pi$ free energies. This also corroborates that the role of the $J/\psi\pi$ is
not essential.

Let us pay attention to the levels placed in the vicinity of the
$D^\ast \bar{D}$ threshold. For simplicity, we first look at the
single elastic channel case.  There appears always a state just below
threshold, as it should occur since we are putting an attractive
interaction in a finite box. As the size of the box increases, and since
there is no bound state in the infinite volume limit (physical case),
this level approaches to threshold.\footnote{This is also discussed in
  more detail in Ref.~\cite{Albaladejo:2013aka}.} When the $J/\psi\pi$
channel is switched on, the $L-$behaviour of this level will be
modified, specially 
when it is close to a discrete $J/\psi\pi$ free energy.   
Note that the slopes of the $J/\psi \pi$ free levels,
in the range of energies considered here, are larger (in absolute value) than those
of the $D\bar{D}^\ast$ ones, because the threshold of the  $J/\psi\pi$
channel is far
from the region studied.

From the above discussion, one realizes that the next coupled channel
energy level, located between the two $D^\ast\bar{D}$ free ones
($E^{\,(0)}_{D^\ast\bar{D}}$ and $E^{\,(1)}_{D^\ast\bar{D}}$), could
be more convenient to extract details of the $Z_c(3900)^\pm$
dynamics. Indeed, in the resonance scenario, this second energy level
is very shifted downwards with respect to
$E^{\,(1)}_{D^\ast\bar{D}}$, since it is  attracted towards the $Z_c$
resonance energy.\footnote{For physical pions ($m_\pi \sim 140\ \text{MeV}$), the $Z_c$
  resonance mass, ignoring errors, is $3894\ \text{MeV}$
  ($3886\ \text{MeV}$) for $\Lambda_2 = 1\ \text{GeV}$
  ($0.5\ \text{GeV}$), as seen from Table~\ref{Table:C-parameters}. For $m_\pi =
  266\ \text{MeV}$ as used in Ref.~\cite{Prelovsek:2014swa}, and
  taking into account the shift in Eq.~\eqref{eq:shift}, one might
  estimate that mass to be around $3912\ \text{MeV}$
  ($3902\ \text{MeV}$).} In this context, it should be noted that the
presence of $Z_c(3900)^\pm$ does not induce the appearance of an
additional energy level, but a sizeable shift of the energy levels
with respect to the non-interacting ones. Therefore, even if no extra
energy level appears, it would not be possible to completely discard
the existence of a physical state (resonance). The energy shift, however, can
be quite large and, only in this sense, one might speak of the
appearance of an additional energy level. The correction of the second energy level in the virtual state scenario is much less pronounced. We should note here that the elastic phase shift computed with the $T$-matrix in Ref.~\cite{Albaladejo:2015lob} does not follow the pattern of a standard Breit-Wigner distribution associated to a narrow resonance. Indeed, the phase shift does not change quickly from $0$ to $\pi$ in the vicinity of the $Z_c(3900)$ mass, and actually it does not even reach $\pi/2$. This is mostly due to a sizeable background in the amplitude.

We now compare the cases $\Lambda_2 = 1\ \text{GeV}$
(Fig.~\ref{fig:energy_levels}) and $\Lambda_2 = 0.5\ \text{GeV}$
(Fig.~\ref{fig:energy_levels-500}). For $\Lambda_2=0.5\ \text{GeV}$,
the relevant (second) energy level is more shifted with respect to
$E^{(1)}_{D^\ast\bar{D}}$ in the resonance scenario
(Fig.~\ref{fig:energy_levels-500}, left) than in the virtual scenario
(Fig.~\ref{fig:energy_levels-500}, right). This is the same behaviour
already discussed for $\Lambda_2 = 1\ \text{GeV}$. However, the shift
for the resonance scenario is smaller in the $\Lambda_2 =
0.5\ \text{GeV}$ case (Fig.~\ref{fig:energy_levels-500}, left) than in the
$\Lambda_2 = 1\ \text{GeV}$ one (Fig.~\ref{fig:energy_levels},
left). This is due to the fact that the $Z_c(3900)^\pm$ is closer to
the threshold and the coupling to $D^\ast\bar{D}$ is smaller for the
$\Lambda_2 = 0.5\ \text{GeV}$ case. Another important difference
between the  $\Lambda_2 = 1\ \text{GeV}$ and $\Lambda_2 =
0.5\ \text{GeV}$ results is that the error band of the relevant energy level is
smaller when the lighter cutoff is used. This is due to the different
relative errors in both cases, and the fact that for $\Lambda_2 =
0.5\ \text{GeV}$, the relevant level is closer to the
$E^{(1)}_{D^\ast\bar{D}}$ free energy than in the $\Lambda_2 =
1\ \text{GeV}$ case.

\begin{figure}[t]\centering
 \includegraphics[height=7.cm,keepaspectratio]{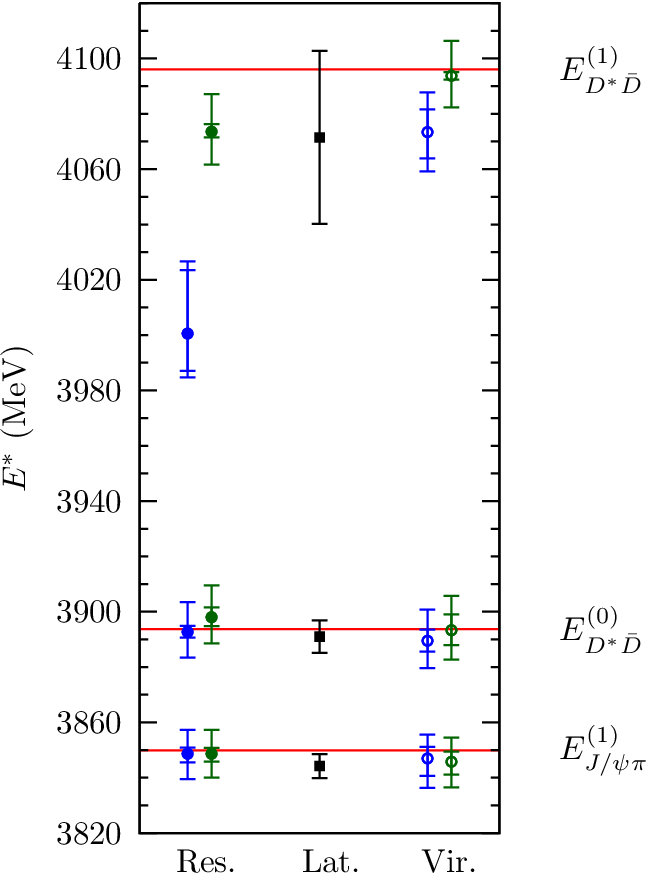} 
 \caption{Comparison of the energy levels of
   Ref.~\cite{Prelovsek:2014swa}, shown with black squares, with our
   results for $L \simeq 2\ \text{fm}$. 
  Full (empty) circles stand for the energy levels obtained in the
   resonance (virtual state) scenario for the $Z_c(3900)$ state. On
   the other hand, the energy levels for the $\Lambda_2 =
   1\ \text{GeV}$ ($0.5\ \text{GeV}$) case are shown by  blue (green)
   circles. The energy levels calculated in this work are displayed
   with two types of error
   bars: the smaller ones have been obtained considering only the errors
   of the parameters entering in the $T$-matrix
   (Table~\ref{Table:C-parameters}), whereas the larger ones
   additionally take into account the errors of the lattice parameters
   (Table~\ref{Table:Lattice-parameters}). \label{Fig:Energy-Levels-zoom}
 }
\end{figure}

After having explored the volume dependence of the energy levels
predicted with our $\tilde{T}$-matrix and scrutinized its physical
meaning, we can now compare our results with those reported in
Ref.~\cite{Prelovsek:2014swa}. The energy levels in the latter work
are obtained from a single volume simulation, $L=1.98 \pm
0.02\ \text{fm}$, and are shown in Fig.~\ref{Fig:Energy-Levels-zoom}
with black squares. In  the figure, we also show the results obtained in this work for $L=2\ \text{fm}$, for
both the resonance (filled circles) and virtual state (empty circles)
scenarios for the $Z_c(3900)$. Besides, the energy
levels  calculated with $\Lambda_2 = 1\ \text{GeV}$ and $\Lambda_2 =
0.5\ \text{GeV}$ are represented in blue and green, respectively. We
provide two different error bars for our results, considering only the
uncertainties    of the parameters entering in the $T$-matrix
(Table~\ref{Table:C-parameters}), or additionally taking into account the errors of the lattice parameters (Table~\ref{Table:Lattice-parameters}). We clearly see  three distinct 
regions, the lowest energies are very close to the $D\bar{D}^\ast$
threshold ($E_{D^\ast \bar{D}}^{(0)}$) and to the first $J/\psi\pi$
free energy level ($E^{(1)}_{J/\psi\pi}$). These free energies are
shown in Fig.~\ref{Fig:Energy-Levels-zoom} with red solid horizontal
lines. As expected, the two lowest lattice levels agree
well with our results for both cutoffs and the two $Z_c(3900)$ state
interpretations  examined in this work. The higher energy levels are the
relevant ones, and, as already mentioned, our results are significantly shifted to lower
energies with respect to $E^{(1)}_{D^\ast\bar{D}}$ for the resonant
scenario, while this shift is much smaller for the virtual state
one. In general, the lattice results are in very good agreement with
the virtual state scenario level for both $\Lambda_2=0.5\ \text{GeV}$
and $\Lambda_2=1\ \text{GeV}$ cases, whereas in the resonance scenario
the agreement is also very good for $\Lambda_2 = 0.5\ \text{GeV}$, and
it is not so good for $\Lambda_2 = 1\ \text{GeV}$. However, in the
latter case, we find  $E_\text{th}=4000^{+24}_{-13}\ \text{MeV}$,
while the lattice energy  is $E_\text{lat} = 4070 \pm
30\ \text{MeV}$~\cite{Prelovsek:2014swa}, and hence this non-compatibility is small, the
difference being $E_\text{lat} - E_\text{th} = 70 \pm
40\ \text{MeV}$. The comparison of our results with those of
Ref.~\cite{Prelovsek:2014swa} support the conclusions given in the
latter work: from the energy levels found in that LQCD simulation one
cannot deduce the existence of a resonance (a truly physical state,
instead of a virtual state), namely $Z_c(3900)$. But also from this
comparison, putting this conclusion in the other way around, one cannot
discard its existence either.

\begin{figure}[!t]\centering
\includegraphics[height=5.1cm,keepaspectratio]{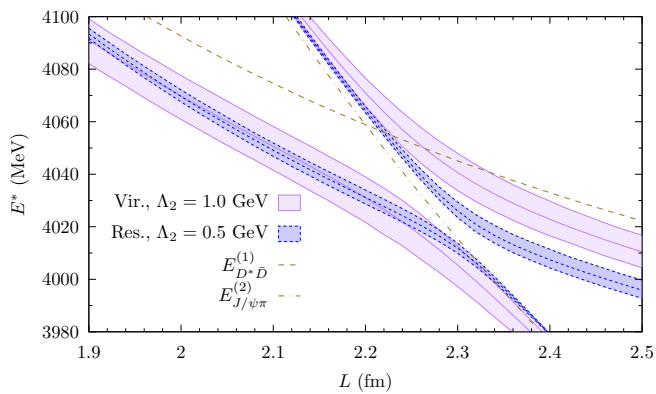}
\caption{Comparison of the relevant energy level for the $\Lambda_2 =
  1\ \text{GeV}$ virtual state (solid purple lines) and
  the $\Lambda_2 = 0.5\ \text{GeV}$ resonance scenarios (dashed
  blue lines)  around $L \simeq 2\ \text{fm}$. The green dashed and dashed-dotted lines represent
  $E_{D^\ast\bar{D}}^{(1)}$ and $E_{J/\psi\pi}^{(2)}$ non-interacting
  energies, respectively.\label{fig:compRV}}
\end{figure}

Finally, as can be seen in Fig.~\ref{Fig:Energy-Levels-zoom}, a
comparison of the relevant energy level obtained in the resonance
scenario for $\Lambda_2 = 0.5\ \text{GeV}$ (green filled circle) with
that obtained in the virtual scenario for $\Lambda_2 = 1\ \text{GeV}$
(blue empty circle) shows that, within theoretical uncertainties (the smallest
error bars), both cases are indistinguishable. This fact can already be seen
by comparing the left panel of Fig.~\ref{fig:energy_levels-500} and
the right panel of Fig.~\ref{fig:energy_levels} around $L \simeq
2\ \text{fm}$. These  energy levels are shown together in
Fig.~\ref{fig:compRV}. It can be seen that, although these two
scenarios cannot be distinguished at $L\simeq 2\ \text{fm}$ (the
volume used in Ref.~\cite{Prelovsek:2014swa}), they lead to appreciably  different
energies already at $L\simeq 2.5\ \text{fm}$. This means that one cannot
elucidate the nature of this intriguing $Z_c(3900)$ state with LQCD
simulations performed in a single volume. Rather, it would be useful
to perform simulations at different values of the box size, to
properly study the volume dependence of the energy levels. Of course,
as discussed in Ref.~\cite{Prelovsek:2014swa}, this would bring in a
technical problem --the appearance of more $J/\psi\pi$ free energy
levels in the energy region of interest, as can be seen in
Fig.~\ref{fig:compRV} ($E_{J/\psi\pi}^{(2)}$). Notwithstanding these difficulties, our work
should stimulate this kind of studies.

\section{Summary}\label{sec:sum}

With the aim of shedding light into the nature of the $Z_c(3900)$ state, we have
implemented  the $J/\psi\pi$, $D^\ast\bar{D}$ coupled
channel $T$-matrix of Ref.~\cite{Albaladejo:2015lob} in a finite
volume, and we have compared our predictions with the results obtained in the LQCD simulation of
Ref.~\cite{Prelovsek:2014swa}. The model of
Ref.~\cite{Albaladejo:2015lob} provides a similar good description of  the experimental information concerning the $Z_c(3900)$
structure in two different scenarios. In the first
one, the $Z_c(3900)$ structure is due to a resonance originating from
the $D^\ast \bar{D}$ interaction, while in the second one it is
produced by the existence of 
a virtual state. We have studied the dependence of the energy levels 
on the size of the finite box for both scenarios. For the volume used
in Ref.~\cite{Prelovsek:2014swa}, our results compare well with the
energy levels obtained in the LQCD simulation of Ref.~\cite{Prelovsek:2014swa}. However, the agreement is similar in
both scenarios (resonant and virtual) and hence it is not possible to
privilege  one over the other. Therefore and in order to clarify
the nature of the  $Z_c(3900)$ state, we suggest performing further LQCD
simulations at different volumes to   study the volume dependence of
the energy levels.

\newpage

\begin{acknowledgements}
We would like to thank S.~Prelovsek for reading the manuscript and for useful discussions. M.~A. acknowledges financial support from the ``Juan de la Cierva'' program
(27-13-463B-731) from the Spanish MINECO. This work is supported in part by the
Spanish MINECO and European FEDER funds under the contracts FIS2014-51948-C2-1-P, FIS2014-57026-REDT and SEV-2014-0398, and by Generalitat
Valenciana under contract PROMETEOII/2014/0068.
\end{acknowledgements}

\bibliographystyle{plain}

%\end{multicols}
\end{document}